# The silicon micro-strip detector plane for the LOFT/Wide Field Monitor

A. Goldwurm [a b 1], P. Ferrando [a b], D. Götz [b c], P. Laurent [a b], F. Lebrun [a b], O. Limousin [b c], S. Basa [d], W. Bertoli [a], E. Delagnes [e], Y. Dolgorouky [a], O. Gevin [e], A. Gros [b c], C. Gouiffes [b c], F. Jeanneau [e], C. Lachaud [a], M. Llored [d], C. Olivetto [a], G. Prévôt [a], D. Renaud [b c], J. Rodriguez [b c], C. Rossin [d], S. Schanne [b c], S. Soldi [b c], P. Varniere [a]

[a] AstroParticule et Cosmologie (APC), 10, rue Alice Domont et Léonie Duquet, 75205 Paris Cedex 13, France; [b] Service d'Astrophysique/IRFU/CEA - Saclay, 91191 Gif sur Yvette, France; [c] Astrophysique Instrumentation et Modélisation (AIM) in SAp/CEA-Saclay; [d] Laboratoire d'Astrophysique de Marseille (LAM), 38 rue F. Joliot-Curie, 13388, Cedex 13 Marseille, France; [e] SEDI/IRFU/CEA - Saclay, 91191 Gif sur Yvette, France.

## ABSTRACT

The main objective of the Wide Field Monitor (WFM) on the LOFT mission is to provide unambiguous detection of the high-energy sources in a large field of view, in order to support science operations of the LOFT primary instrument, the LAD. The monitor will also provide by itself a large number of results on the timing and spectral behavior of hundreds of galactic compact objects, Active Galactic Nuclei and Gamma-Ray Bursts. The WFM is based on the coded aperture concept where a position sensitive detector records the shadow of a mask projected by the celestial sources. The proposed WFM detector plane, based on Double Sided micro-Strip Silicon Detectors (DSSD), will allow proper 2-dimensional recording of the projected shadows. Indeed the positioning of the photon interaction in the detector with equivalent fine resolution in both directions insures the best imaging capability compatible with the allocated budgets for this telescope on LOFT. We will describe here the overall configuration of this 2D-WFM and the design and characteristics of the DSSD detector plane including its imaging and spectral performances. We will also present a number of simulated results discussing the advantages that this configuration offers to LOFT. A DSSD-based WFM will in particular reduce significantly the source confusion experienced by the WFM in crowded regions of the sky like the Galactic Center and will in general increase the observatory science capability of the mission.

## 1. INTRODUCTION

The Wide Field Monitor (WFM) is the second instrument of the Large Area Observatory for X-ray Timing (LOFT) mission [7]. Its main objective is to provide unambiguous detection and localization of the high-energy sources in a large field of view, in order to support science operations of the LOFT primary instrument, the Large Area Detector (LAD). The WFM is a coded aperture telescope where a position sensitive detector records the shadow of a mask projected by the celestial sources. Specific data processing of the recorded detector images (usually denoted as "deconvolution") allows one to reconstruct the sky image generally over a very large field of view. Indeed coded mask telescopes are nowadays the best instruments able to provide, in the frequency band of the X and the hard X-rays, reasonable angular resolutions and sensitivities over very large fields of view.

The original design of the WFM is based on Silicon Drift Detectors (SDD), the same sensors used for the LAD. Even though SDD provide very good spectral performances their positional response is asymmetric and their use for an imaging system implies a combination of two orthogonal systems with a resulting point spread function with cross-shaped side-lobes. Since sensitivity, telemetry allocation and operating modes foreseen for the WFM do not allow employing fully the spectral capabilities of the SDD, an alternative option for the WFM detector plane is proposed, which improves the imaging performances and maintain the spectral ones at a level for which no science losses are expected.

---

[1] andrea.goldwurm@cea.fr; phone +33169088669 or +33157278058.

This alternative design for the WFM detector plane is based on Silicon Double Sided Strip Detector (DSSD) sensors that allow proper 2-dimensional imaging. By having the incoming photons positioned in both directions with equivalent fine resolution, the proposed instrument makes of the WFM a really full-2D imager and therefore provides the best imaging capabilities compatible with the allocated budgets for the WFM on LOFT. This paper describes the proposed 2D-WFM option and in particular the detector plane and the associated electronics that are envisaged for this configuration. We will also present the global performances of both detector plane and the overall WFM instrument for the 2D detector option, through calculations, simulations and tests. While reducing the complexity of the overall instrument, the proposed configuration will significantly improve the imaging capabilities of the system since it will provide an effective resolution of $< 5'$ along all directions (instead than the $5' \times 8°$ plus $8° \times 5'$ resolutions obtained with the original 1.5D system). The reconstructed images will be much improved because of the absence of the secondary lobes of the PSF along the instrument axis and the reduced coding noise along the whole sky reconstructed image, again due to the much smaller mask element dimensions.

This will increase the WFM scientific performances for a number of science cases by
- providing an unambiguous localisation of transient events, wherever they will appear in the WFM FOV, as far as they are distant from another source by $> 3' – 5'$ in any direction
- avoiding confusion and providing source localisation even in the dense Galactic Center region,
- reducing the error on source flux measurements due to contamination by other close-by sources,
- allowing for search of weak sources in summed up images, leading to a deep hard X-ray survey at the end of the mission at the 100 microCrab level
- increasing the grasp of the WFM for the same number of cameras.

## 2. INSTRUMENT DESCRIPTION

### 2.1 Overall WFM configuration

The 2D WFM system is composed by a set of identical coded mask telescopes or cameras that are mounted on a mechanical structure on the top of the LOFT optical bench. Each camera is a standalone system that is itself composed by a mechanical structure, a 2D coded mask, a collimator, other subsystems, and a detector housing box containing: the detector plane with its Front End Electronics (FEE) and in particular its ASICs, the Control and Read-out Unit (CRU), the Power Supply Unit (PSU) for this camera. The baseline for the overall WFM configuration is made of 4 cameras disposed along the direction parallel to the solar panels to offer a total FOV that covers as much as possible the region of sky accessible by the LAD, as shown in Figure 1 (left). The design of a single camera (Figure 1 right) follows the classic design of a 2D coded mask system.

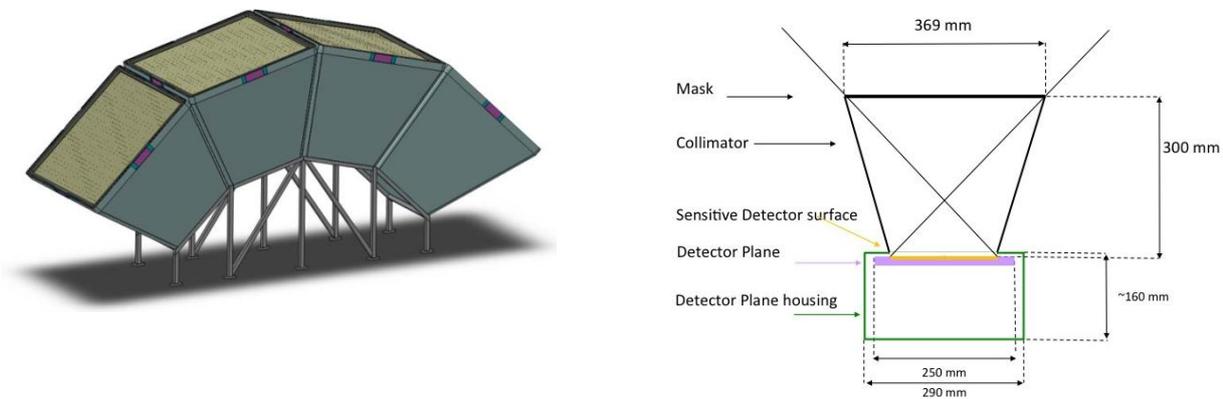

Figure 1 Left: baseline configuration for the WFM instrument; right: sketch of each individual camera.

The mask is positioned at a distance of 300 mm from a position sensitive detector plane and a collimator between the mask and the detector protects the sensor from particles and radiation coming from outside the Field of View. The 2D, 0.150 mm thick, square coded mask of tungsten of total dimensions of 369 × 369 mm², is composed by identical basic square elements of dimension 0.360 × 0.360 mm², and has an open fraction of 0.25, which optimises the signal to noise (SNR) for weak sources in presence of large celestial background. The pattern is random, in order to minimize and distribute in uniform way the coding noise produced by the sources in the field. These dimensions can indeed be further optimised depending on which parameters (angular resolution, field of view) have to be improved. The flexibility of the proposed system will allow us to easily change this design without significantly modifying the overall budget and external interfaces, then allowing optimisation of the number of cameras and their disposition.

### 2.2 Detector Box, plane and module

The detector plane is hosted in a detector box (detector plane housing in Figure 1) which is organised in three different levels: the detector case, hosting the detector plane itself with detectors and FEE, the level hosting the Control and Read-out Unit (CRU), sequencers and ADCs, and the level hosting the Power Supply Unit (PSU). Each of these levels has a specific interface with the rest of the WFM camera or the spacecraft. The detector plane has a thermal interface, through a cold finger, with the regulation system provided by the S/C. The CRU has a space-wire interface with the BEE. The PSU interfaces with the S/C power lines. The estimated dimensions of the detector housing box are 290 mm×290 mm×160 mm. A possible implementation of the system, as three separate cases, is shown in Figure 2.

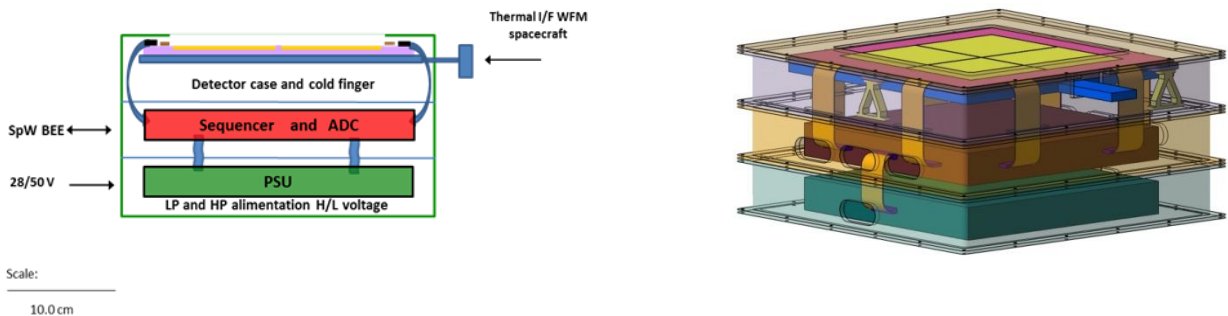

Figure 2: Possible design scheme (left) and mechanical drawing (right) of the detector housing box with a concept of three independent cases.

The detector plane is composed of 4 detector modules assembled on a mechanical structure as shown in Figure 2. This supporting structure includes a metal support, a cold plate connected to the cold finger and thermal washers to isolate it from the base of the detector case. The total dimensions of the detector plane (not including the washers and the flexi connectors) are: 250 mm×250 mm×20 mm. The flexi connectors between the FEE and the CRU in the case below have an estimated minimum curvature radius of 11 mm. This fits easily in the 29 cm size of the box mentioned above.

The detector plane imaging area is a square of size 189 mm×189 mm (357.2 cm²), with a central inner cross of 5 mm width not active (guard ring and mechanical mounting, see Fig. 3), representing 18.6 cm² of dead area. A detector module (four per detector plane) is made of a single silicon Double-Sided Strip Detector, with an overall area of 97×97 mm² (sensitive area 92×92 mm²), and a thickness of 500 μm. On each DSSD face are located 512 micro-strips with a pitch of 180μm. The orientation of the strips on one side is at 90° from the orientation on the other side. The effective live geometrical area of the detector plane is therefore 4 × 92 mm × 92 mm (386.6 cm²) and it is viewed by 4096 (2 × 2 × 512 per side) channels. The DSSD is mounted on a ceramic carrier, 2 mm thick, as shown in Figure 3. The ceramic is about three centimetres larger than the DSSD, in order to accommodate the FEE elements, the ASICs and associated passive parts. The DSSDs strips are connected to ASICs (64 channels each) through a pitch adapter. There are a total of 16 ASICs on the ceramic, 8 per DSSD side. The module is connected to the CRU, in the second level of the detector housing box, through a flexible connector, with a length of less than ~ 30 cm.

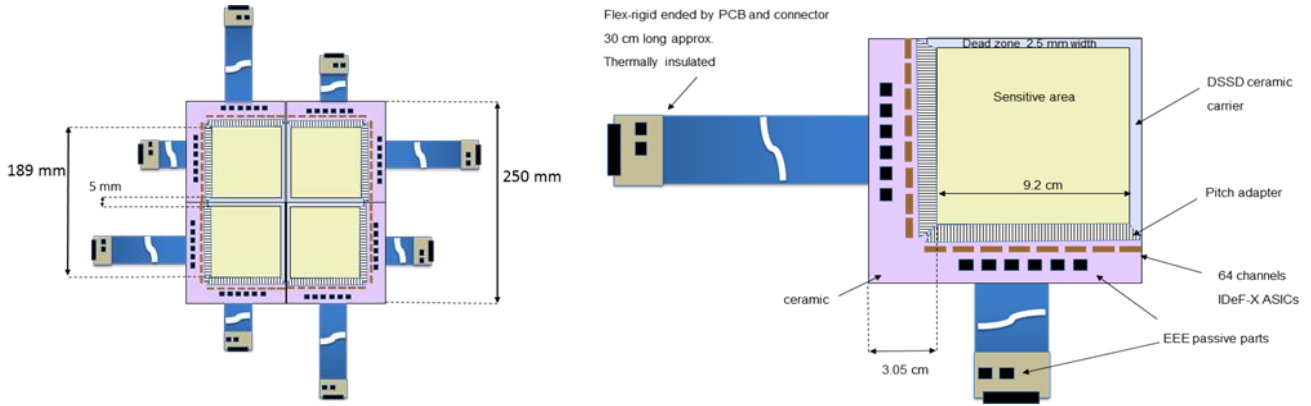

Figure 3: Detector plane assembly (left) and module design (right).

### 2.3 Detector and front end electronics

The chosen detectors for the proposed instrument, the Double-Sided Strip Detector (DSSD), have been widely used as trackers in particle physics experiments and have been (AMS) or will be soon (Astro-H) used in space, as X-ray spectro-imagers in the latter case. They provide a robust and efficient 2D-imaging system [23]. The spectral performance of the detector system depends strongly on the DSSD design (capacitance, leakage current), the DSSD-ASIC coupling and last but not least, the ASIC performance. The Idef-X ASIC and its variants developed since 10 years are ideally suited for the readout of these detectors.

The DSSD is produced from a silicon n-type square wafer with sensitive area of 92 mm × 92 mm and each side covered by 512 strips with a pitch of 180 µm surrounded by a guard ring. The strips are oriented at 90° on one side with respect to the other one, in order to provide the two coordinates of the interaction point in the detector. The DSSD is a p-n-n$^+$ junction where electrodes are obtained by implantation of the doping material into an n-type silicon wafer. The side with the implantation of a p-type material is called the junction side and the other side the ohmic side. The detector thickness is 500 µm. This is a standard for the industry, and it provides a good efficiency in the required energy range. The DSSD pitch impacts on many important performance parameters but to minimize the differences between strips and the external capacitance, it is best to have a close matching of the DSSD strip pitch with the ASIC channels pitch. We have therefore chosen a strip pitch of 180 µm in order to nearly match that of the Idef-X ASIC channels (150µm) and at the same time provide sufficient space to dispose the ASICs along the detector module external side. This choice allows us to propose a version of the Idef-X ASICs that is available and has been already evaluated at a very high TRL.

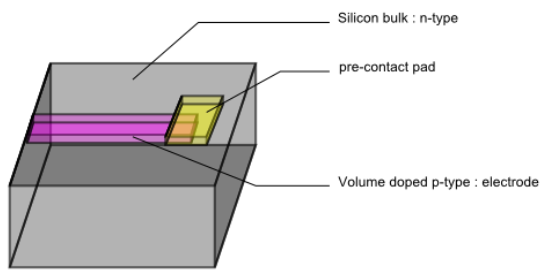

Figure 4a. Outline of an electrode on the junction side. This side, except for the polarization pads, is coated by a passivation layer with a thickness of ~ 100 nm (not shown on the figure).

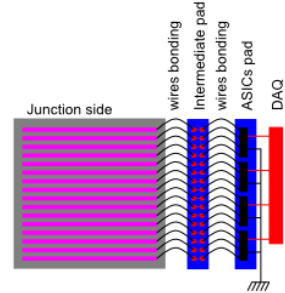

Figure 4b. Sketch of the DC configuration on the junction side. Each of the 512 strips is connected to one of the 8 ASICs devoted to this side.

One important issue for DSSDs is their bulk and inter-strip capacitance. To first order, the former is proportional to the strip pitch and inversely proportional to the DSSD thickness and the latter is inversely proportional to the inter-strip gap. For a fixed thickness and width-to-pitch ratio, the total detector capacitance is almost independent of the pitch. Conversely, for a fixed thickness and pitch, both the body and inter-strip capacitances increase with the strip width. On the other hand, the metallic deposition of the strips (aluminium) absorbs the low energy X-rays and detectors with very small strip width are more sensitive to radiation. With these considerations in mind, we have considered a reasonably small strip width of 30 μm and a large gap of 150 μm.

The junction side will be connected in a DC-configuration, which allows having a very thin entrance window on this p-side, as shown in Figure 4. This will let low energy photons interact within the effective detector volume. This, together with the better energy resolution than on then n-side (see e.g. [27]), will allow to match the sensitivity requirement at 2 keV, by exposing this side to the sky. In this configuration, the polarization of the strips, and their signal measurement, are both insured by the ASICs Idef-X. The implementation, with an intermediate plate, is shown in Figure 4. Besides its role as a pitch adaptor, this intermediate plate also allows to perform test measurements of the DSSD without the ASICs.

The ohmic side is usually less efficient on the energy resolution and position measurement than the junction side due to its architecture. Indeed, its design is more complicated, as shown in Figure 5a. For example, in order to decrease the charge trapping on the edge this side is equipped with p-type implanted electrode, so-called p-stop, between the n-type electrodes. Each of the 512 n-type electrodes is polarized by a bias voltage (~ 100 V) through a deposited resistor of at least 500 MΩ. In order to work with an ASIC with a grounded reference voltage, it is preferable to have on the DSSD a capacitor decoupling function with a capacitance value of at least 150 pF. This capacitor is obtained by depositing a dielectric layer coated by a metallic deposited layer (Fig. 5a). The Figure 5b shows the equivalent circuit and the connection to the ASICs.

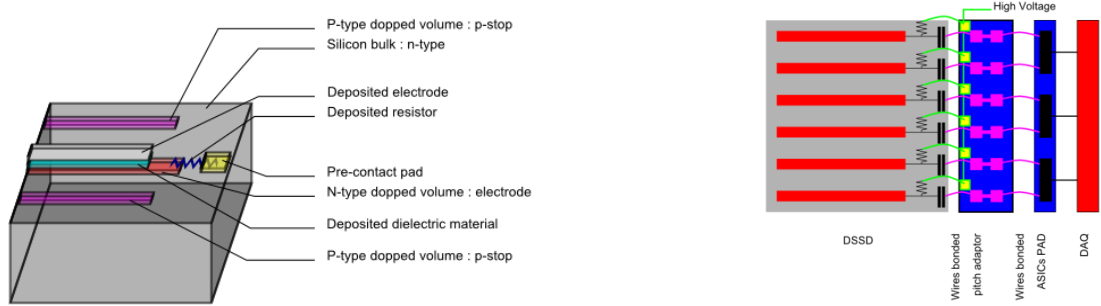

Figure 5a. Outline of an electrode on the n-side. As for the p-side, there is also a ~ 100 nm passive layer not shown on this figure.

Figure 5b. Sketch of the AC configuration on the p-side. There are 8 ASICs for 512 strips on this side.

The estimated values of the leakage current per strip is 7 to 10 pA at -20° C for both Ohmic and Junction sides and of the total capacitance per strip is 10 to 12 pF and 7 to 10 pF for Ohmic and Junction sides respectively. These values are fully consistent, or even conservative, with respect to existing measurements on similar devices, as discussed below.

According to DSSD manufacturer and devices measurements, the typical current density is expected to be in the range from 20-30 pA/mm$^2$ at room temperature when a 500 μm thick sample is fully depleted. The surface to be considered per strip is defined by the pitch times the strip length i.e. 16.56 mm$^2$. Thus, the current in a strip is expected to be in the range from 300-500 pA at room temperature. Assuming an activation-energy of 0.62 eV, the expected current at –20°C is in the range 7-10 pA/strip. In our performance evaluation, we consider a 10 pA current as a worst case at -20°C.

Regarding the total capacitance per strip seen by the ASIC, we consider that our DSSD junction side value will be augmented by 3 pF, a typical value for accounting for bonding and interconnection. This total capacitance (thus of 10 to 13 pF/strip) is compared in Figure 6 with other measurements and with a model. The model calculates the bulk and inter-strip capacitance, using the Cattaneo model (a mathematical transformation model) [5] on one hand, and a 2D Poisson solver (an idealized field calculator using over-relaxation method) developed by some of us on the other hand. Both methods give comparable results for low values of strip width / strip pitch. It must be noted that such models represent the ideal case stray capacitance, the minimum value that can be reached in such devices.

The curve plotted in the figure 6 is our model, very weakly dependent on detector thickness, and to which has been added the 3 pF value mentioned above. The model fits well the data at least up to width/pitch values of 0.5, well above our needs. As can be seen, our DSSD value is conservative with existing measurements, and with respect to the model.

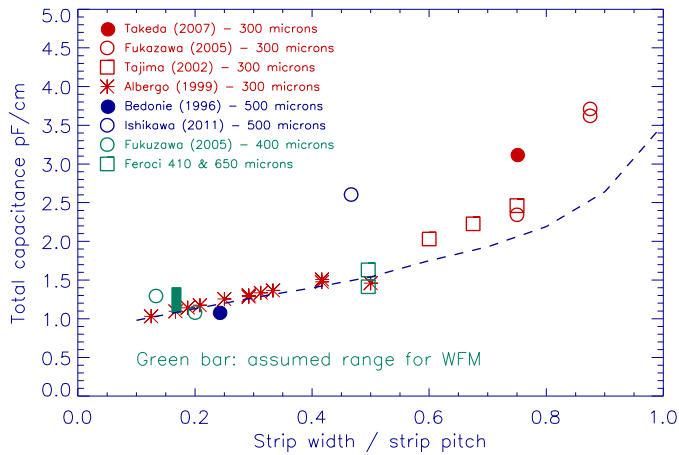
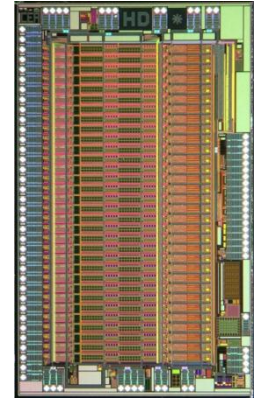

Figure 6. Total capacitance per cm and per strip measured, and compared to the model discussed in text. The green bar is the capacitance we assume for our device.

Figure 7. Picture of IDeF-X HD, with 32 channels. Inputs are located at the left, distributed every 150μm.

The front-end readout chip to be used with DSSD silicon imaging spectrometers into the WFM-2D are ASICs named IDeF-X HD/BD, standing for "Imaging Detector Front-end in hard X-ray" version "HD/BD", signing large dynamic from few keV to hundreds of keV and bi-directionality of input signals, i.e. capability to readout indifferently anodes or cathodes of semiconductor detectors. The circuits, derived from a long R&D program, are full custom ASICs developed at CEA/Saclay that combine the most recent developments for space qualified devices having similar energy range and noise requirements or better than WFM-2D DSSD detectors (see [16,9,10,22,20,21]).

IDeF-X HD/BD is developed from the existing and qualified ASICs, namely IDeF-X BD and IDeF-X HD and analogue building blocks of IDeF-X V0. It will use the same electrical schemes as IDeF-X HD, having the option of reading out either cathode or anode (programmable by slow control) as implemented in IDeF-X BD and optionally using a "large capacitance" Charge Sensitive Amplifier (CSA) if required as implemented in IDeF-X V0. All these circuits are built using the same AMS CMOS 0.35 μm technology and using the same qualified full custom rad-hard proven libraries and design rules. IDeF-X HD/BD is a minor evolution of the two previous versions of IDeF-X to match the required fine pitch geometry and to adapt the front-end preamp to match the DSSD strips and hybrid stray capacitance. IDeF-X HD/BD is equipped with 64 independent spectrometry channels. The channels are grouped into two sets of 32 channels, making this ASIC somehow the sum of two IDeF-HD onto the same substrate. The two chips are considered completely independent from the sequencer but the power supplies are brought together, saving precious surface to match space qualified bonding requirements.

Each individual channel is made of: a CSA optimized for low current (<1nA) and capacitance around 10 pF; a variable gain (inverting or non-inverting) stage; a pole-zero cancellation stage; an adjustable shaper in the range from 1 to 13 μs (peaking time); a baseline holder, providing a stable offset whatever the leakage current into a channel; a peak detector and hold; a discriminator (each channel, 6 bit DAC) in the range from 0 to 13 keV for Si (the settings is not linear to allow fine tuning in the low range and coarse in the high range). The CSA includes a continuous reset feedback circuitry which includes a "non-stationary noise suppressor" to optimize the noise response into the whole dynamic range. The dynamic range is 10 fC (but is programmable up to 40 fC).

When a particle hits the detector, the charge cloud moves along the electrical field lines into the sensor generating a transient current charging the feedback capacitance of the CSA. The signal out of the CSA is filtered and shaped according to the settings (all channels have the same shaper peaking time). The amplified signal is presented to the discriminator circuitry and generates a trigger if the signal level is higher than the threshold programmed into the channel. The trigger is memorized into a hit register as several channels may be hit simultaneously. A global logical OR of 32 channels is sent to the readout sequencer, which will freeze the chip entirely (no change in the peak detector is admitted after this moment) and start the readout sequence according to the selected mode (automatic, on demand, full). As mentioned before, 32 channels are grouped together. The IDeF-X HD is shown in Figure 7. The interface of IDeF-X HD/BD is 64 inputs (DC or AC coupled), 2 differential analogue output (AOUT) and 2 independent slow controls (using LVDS – STROBE, DIN, DOUT, TRIG). Note that the IDeF-X HD/BD is equipped with an on-chip thermal sensor which is readout as a virtual 65$^{th}$ channel. Optionally, two sensors could be placed on the chip. The thermal sensor is well suited and linear in the range from -50°C up to +30°C.

The application or user may choose different readout mode and several operation modes can be provided. The trigger time stamp has a typical accuracy of 100 ns or better. Without any time-walk correction, the accuracy is limited to the peaking time value, which is considered to be 10 µs in this estimation. The IDeF-X ASICs, HD, BD and V0 in particular, have been extensively studied with respect to radiation tolerance. The expected performances applicable to the envisaged version for the 2DWFM are as follow: Radiation SEL free, Radiation SEE > 9 MeV cm$^2$ mg$^{-1}$, Radiation TID > 300 krad w/o effect on noise response. The Single Event Upset threshold value is counterbalanced by the SEU signal, which is a logical flag alerting the controller that a SEU has probably occurred. The SEU management is therefore achievable by monitoring SEU and/or periodically reprogramming the chip. TID has been experienced up to 1 Mrad on certain devices.

### 2.4 The DSSD-ASIC chain performance assessment

The IDeF-X HD/BD has performances identical or better to the current IDeF-X HD version. Appreciating the low noise performance of IDeF-X requires measuring the noise according to the peaking time, choosing various conditions of current and stray capacitance in the chip. Measurement results were obtained with IDeF-X HD when the detector current changes (no additional capacitance at the input, 1pF approximately) and when the stray capacitance changes (fixed circulating current in a DC mode). From this set of measurements, one can derive the noise parameters and double check the simulation accuracy. This allows to precisely estimating the expected noise for given DSSD parameters and optimization of the CSA capacitance matching. Assuming a stray capacitance in a range of 10 to 13 pF per strip and a dark current of 10 pA/strip at –20°C as discussed above, taking into account (simulation) a CSA input transistor matching, we calculate the noise on the P side of a DSSD with a DC coupling (Figure 8) to be in the range of 81-91 el rms, as tabulated below.

Assuming 3.64 eV/electron-hole pair in Si DSSD, 90 el. rms corresponds to a threshold of 2 keV at 6 sigmas and an energy resolution of 800 eV FWHM at 6 keV, for 13 pF of capacitance. On the N side, assuming an extra capacitance of 20 %, the same current and an AC coupling to the chip, the noise is expected to be in the range of 125 el. rms. This corresponds to a 2 keV threshold at 4.4 sigma (for 13 pF). We do not intend to use the N side as a spectrometric channel but as a trigger only. We have detailed above the worst case parameters. The effective threshold, at 6 sigmas, for both sides are given in Table 1, for the range of leakage current and stray capacitance given in the DSSD section. On the P-side, a threshold as good as 1.7 keV can be obtained. In calculating this threshold, one should finally consider that in IDeF-X designs, the noise on the discriminator path is slightly lower than the noise (ENC) on the energy path. Typically, the discriminator can be 5-10% better. To keep conservative, we used only the ENC on energy path to derive the performances of DSSD threshold. This remark does not affect the energy resolution.

As noted above, the N-side has a higher noise than the P-side, 125 e$^-$ rms with respect to 90 e$^-$ rms (for 13 pF). The threshold will obviously be set at the same value, 2 keV, for both sides, so that the N-side will have a larger trigger rate on noise than the P-side. The good N-side triggers, corresponding to a photon, are accompanied by a trigger in coincidence on the P-side. Requiring such a coincidence, in a time window equal to the peaking time, ~ 10 µs, will allow rejecting the "noise" N-side triggers with a high efficiency. The question remains however of possible incorrect N-side position assignments, due to the possible presence in a coincidence window of pure noise events because of the 4.4 sigma threshold. The probability of such a wrong N-side position assignment has been calculated, and is at most 0.5 % at the threshold energy of 2 keV. This probability decreases rapidly with energy. We thus see that the system is indeed very robust for the imaging capabilities, starting from the threshold energy, even if the N-side has a larger noise than the P-side.

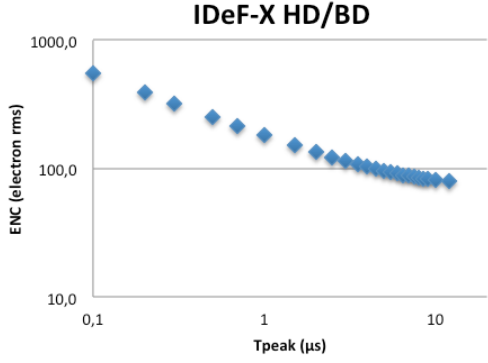

Figure 8. Predicted noise performance of IDeF-X HD/BD, assuming a 10 pA current in a strip, 12 pF stray capacitance, on the P side of a DSSD, DC coupled to the input. The CSA is matching the stray capacitance at the input.

Table 1. Threshold at 6 sigmas (LT) on P and N-sides as a function of the stray capacitance C and of the leakage current I.

| Side | C (pF) | I (pA) | R (MOhm) | ENC (el. rms) | LT (keV) | En. Res. at 6 keV (eV) |
|---|---|---|---|---|---|---|
| P | 10 | 5 | / | 78 | 1,7 | 678 |
| P | 10 | 10 | / | 81 | 1,8 | 698 |
| N | 10 | 5 | 500 | 115 | 2,5 | 983 |
| N | 10 | 10 | 500 | 117 | 2,6 | 1001 |
| P | 13 | 5 | / | 88 | 1,9 | 752 |
| P | 13 | 10 | / | 91 | 2,0 | 778 |
| N | 13 | 5 | 500 | 124 | 2,7 | 1061 |
| N | 13 | 10 | 500 | 125 | 2,7 | 1069 |

## 3. INSTRUMENT PERFORMANCES

### 3.1 Imaging performances

The proposed 2DWFM imaging system is based on the coded mask telescope principles. For well-defined 2D coded mask systems, estimates of the expected performances can be given (see [25]) from the system configuration. However these estimates are valid for an on-axis isolated source and rely on the assumption of a system totally free from coding noise and with a side-lobes free Point Spread Function (PSF). We have therefore run detailed imaging simulations of the system and analysed the simulated data by using standard reconstruction and analysis algorithms already employed and tested for other coded mask missions [12, 13] in order to verify the predicted performances in realistic conditions.

The spatial response of DSSD is approximated by a 2D box function of width equal to the strip pitch (= 180 μm in our case) independent of energy, equivalent to the one provided by a pixelised detector. The equivalent FWHM spatial resolution is a factor $[2.35 / (3\sqrt{2})]$ of the box width, which gives in our case a value of 122 μm. The events are therefore collected in images where a pixel corresponds in size to the strip pitch sizes and is centred at the intersection of the strips. Given the camera configuration described above, the imaging system will be characterized by the following parameters: recorded images of 1050×1050 detector pixels out of which 1024×1024 sensitive (the dead area is a cross of 26 pix width in the center of the image), a mask of 1025×1025 elements with a ratio of 2×2 detector pixels per mask element, reconstructed sky images of 3100×3100 pixels with on-axis sky pixel angular size of 2.06' and projected mask element angular size (that represent the geometrical FWHM angular resolution) of 4.1'.

Different simulations have been performed, from isolated on and off-axis sources to complex distributions of hundreds of sources over the FOV as those seen in the Galactic Bulge and including a cosmic X-ray background as given in [1]. In particular several simulation runs of the galactic bulge using the 4[th] INTEGRAL/IBIS catalogue [4] for the source parameters, with different exposures from 10 to 200 ks, different energy ranges and pointing directions have been performed. These simulations have allowed us to study the basic characteristics of the 2D-WFM system and in particular the extent of the field of view, the location, shape and width of the main peak of the point spread function (PSF) and the extent and influence that its secondary lobes produce on the image (coding noise) and on the measurement of the characteristics of the other sources.

The reconstruction algorithm that provides a sky image of the FOV is based on a kind of balanced cross-correlation deconvolution from the simulated detector image and a decoding array derived from the mask pattern (see e.g. [12]) with a normalization that gives a zero average for an empty source sky observation and renormalized intensities to the corresponding on axis value. These operations can be efficiently computed through a set of Digital Fast Fourier Transforms and provide in output four sky images for Intensity, Variance, Exposure and Signal to Noise ratio image.

The field of view at zero response in coded aperture systems is primarily given by the level of modulation induced by the mask on the source radiation. A complete modulation is obtained in the so called Fully Coded FOV and zero modulation is reached at the limit of the partially coded FOV (PCFOV), the FOV external to the FCFOV, at the so called zero response FOV (ZRFOV). Total angular dimensions of these FOVs depend on the mask dimension, the detector dimension, the mask to detector distance. However two other effects shall be considered to compute the effective FOV of a given system. There is of course a reduction of effective flux seen by the detector for a non-axis source due to the inclination of the parallel rays with respect to the detector which is proportional to the cosine of the off-axis angle. The so called "vignetting" is due to the non-zero mask thickness that produces a shrinking of the open mask elements as seen by the detector. It actually depends on the actual disposition of elements of the mask. Combining modulation, vignetting and cos-theta effects, provides the expected FOV of the system as shown in Figure 9, which gives the profile of the effective FOV along one axis of the instrument passing through the telescope axis, with the contribution of the three components. One can see that the FOV at 50% sensitive area is 50° wide and it reaches 88° at zero response, while the FCFOV is 33°. Simulations provide the expected FOV as show in Figure 10.

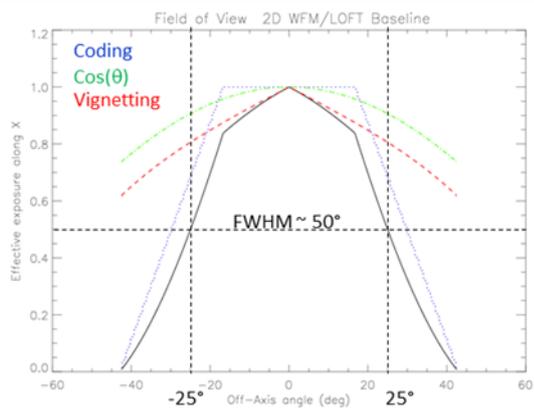

Figure 9. Field of View of one 2D WFM Camera along one axis for the baseline configuration

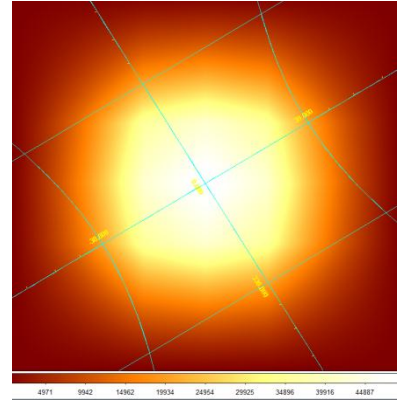

Figure 10. Exposure image of a single camera observation of 50 ks over the whole FOV for the camera configuration.

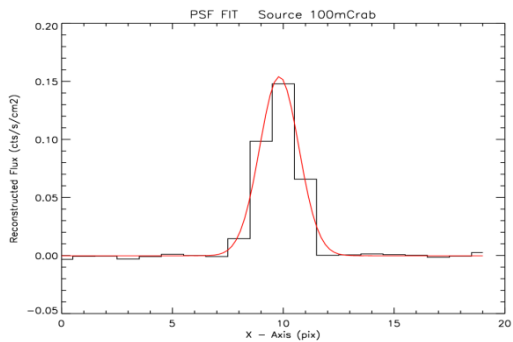

Figure 11. PSF profile of a 100 mCrabs source (SNR = 100) on axis in 1 ks obs. simulation. The best fit Gaussian (solid line) FWHM is 4.2' and the location error is 2.7″.

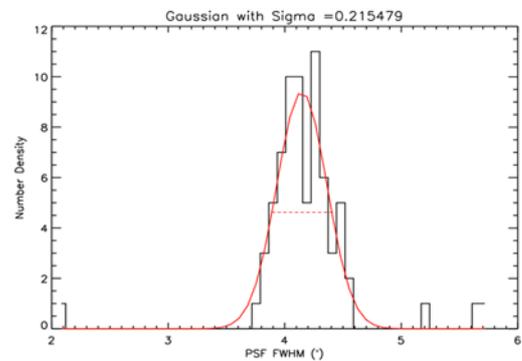

Figure 12. Distributions of PSF fitted widths (FWHM) of sources from a 50 ks GC obs. simulation. The best fit Gaussian (solid line) is at 4.15' and has a sigma of 0.22'.

The angular resolution of a coded mask system is given by Δθ = arctg ($D_{me}$/H) where $D_{me}$ is the mask element size and H the mask to detector distance. For the chosen configuration this gives an angular resolution (FWHM) of 4.13 arcmin on axis for the baseline and upgraded configurations respectively. However a number of effects (tangential projection, detector spatial resolution, vignetting, coding noise side-lobes, etc.) can induce a significant distortion of the PSF in particular for off-axis partially coded sources. The imaging simulations we carried out confirm the expected value of the resolution all over the FOV. Fitting the PSF (Fig. 11) of the detected 70 sources (> 7 sigma) in reconstructed images of a 50 ks GC obs. simulation we inferred an average Angular Resolution of 4.15' ± 0.22' (Fig. 12).

The point source location error (PSLE) in coded mask systems is approximately given by PSLE ~ Δθ / SNR where SNR is the Signal to Noise Ratio of the source and Δθ the angular resolution. For the proposed systems it is therefore expected to be in the range 0.5' – 1.0' for a 10 sigma source. However given the complex interplay between different effects the best way to estimate the PSLE is again through imaging simulations. The 50 ks GC obs. simulation allowed us to fit the PSF of over 70 sources (> 7 σ) at different SNRs and plot the difference between fitted positions and catalogue positions. The PSLE is then plotted versus the source Signal to Noise Ratio (Fig. 14) and the 1/SNR dependence considered above is fitted to the data to obtain the normalization for 90% confidence level. The fitted 90% c.l. PSLE is 40" at SNR=10.

The PSLE depends somehow on the systematic noise. The above values were obtained from images cleaned by the coding noise (Fig. 13), which is dominated, in the GC region, by Sco X-1. The location accuracy in case of the uncleaned images is reduced, but still reaches the values of the WFM scientific requirements (better than 1' for SNR > 20). For observations without Sco X-1 in the FOV, accuracies of the order of 30" at SNR=10 are obtained.

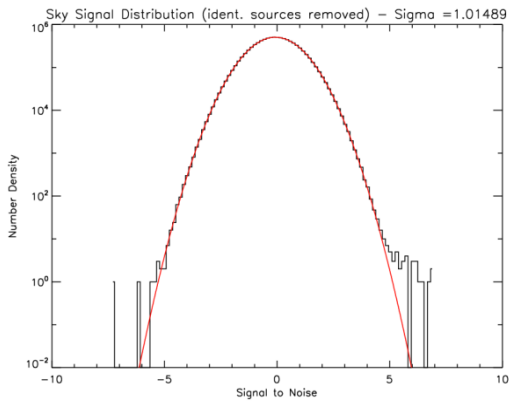

Figure 13. Distribution of excesses in the cleaned SNR image of the 50 ks simulation of the GC, and its best fit Gaussian of standard deviation = 1.01. Sources were cleaned up to SNR of 7.

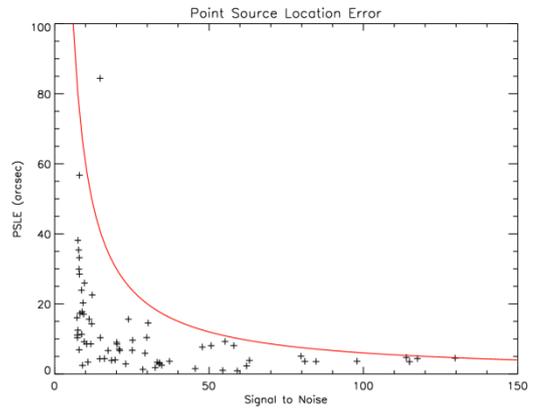

Figure 14. PSLE versus the SNR in cleaned image of the 5 ks GC obs. simulation with the 1/SNR curve normalized at PSLE=60″ at SNR=10. The only point above the curve is from the central source IGR J1745.6-2900, actually a blend of close sources for which the single PSF fit is not adapted.

The capability of the 2D WFM to image complex regions of the sky, where large source density may lead to large coding noise in the images and problems of source confusion, has be tested with imaging simulations of the Galactic Center region. Fig. 16 shows a deep zoom in the center of the reconstructed and cleaned image from a 50 ks observation simulation of the Galactic Center in the 2-10 keV range. It can be seen that the system fully disentangles the contribution of the different sources even in the central half degree of the Galaxy.

### 3.1 Energy response and sensitivity

The energy range and energy resolution of a WFM camera correspond to the energy range and energy resolution of the detectors. No further degradation is expected from the imaging system and data processing apart from that due to binning of data because of the limited TM rate allocated to WFM data transmission. This last one is indeed very important and will limit the effective spectral resolution of the system since it is expected to transmit data in 8 to 16 energy channels.

The energy response of one WFM camera was computed from Monte-Carlo simulations, using the Geant-4 software. The detector geometry and the source properties are parameterized and used as input for the simulation software. The results of the simulation are used jointly with the known energy and position resolution of detectors, to produce realistic data. Coincidence and/or anticoincidence logics are also taken into account at this stage. These data are used afterward to create spectra at different energies, in order to generate the response matrices of the proposed instrument. The absorption effects due to the other materials in front of the detector (e.g. thermal blancket) were then included in effective area.

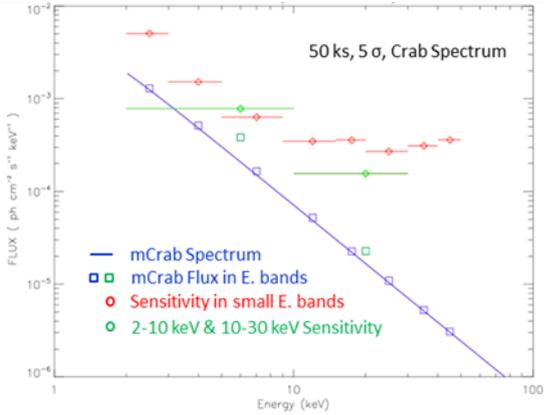

Figure 15. WFM camera Sensitivity on axis in several energy bands and in the 2-10 keV and 10-30 keV ranges for a 50 ks observation at the 5σ significance level. Sensitivity is compared to the mCrab flux density (solid line) and the mCrab fluxes integrated in the same energy intervals (squares).

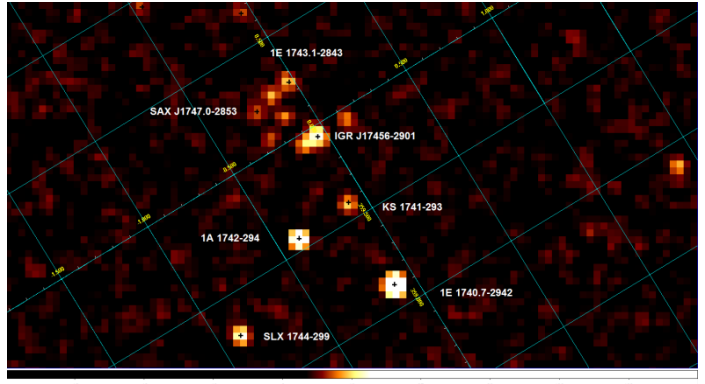

Figure 16. Reconstructed SNR image of a 50 ks observation of the Galactic Center with one WFM camera in the range 2-10 keV and in the 2°×2° inner central region where positions (back crosses) and names of some of the simulated sources are indicated. Note that at the very center the IGR J17456-2901 source is surrounded by 2 close sources, one of which is in fact clearly disentangled.

From the energy response then we computed the sensitivity of the system that is reported in Figure 175. In the energy ranges 2-10, 10-30 and 30-50 keV we obtained respectively 2.0, 6.9 and 53 mCrab (5σ, 50 ks). Imaging simulations have demonstrated that the sensitivity reached in the images is fully compatible all over the FOV with the predicted values.

## 4. CONCLUSIONS: REQUIREMENTS AND PREDICTED PERFORMANCES

The predicted performances of the proposed 2DWFM system for the LOFT mission are summarized and compared to the main WFM requirements in table 2. An upgraded version of the system that increases the FOV (and therefore the grasp) and improves the angular resolution for basically the same sensitivity on axis has also been studied and can be implemented if larger budget of volume, mass and power will be available for the WFM system (Soyouz launcher). Note that the condition for the location accuracy has been converted to source significance in order to directly compare with our PSLE curves. A 10 mCrab source would have in a 50 ks observation a significance of 25 σ if the goal on sensitivity (2 mCrab at 5σ) is reached, as is the case for the system we propose.

As shown above the energy low threshold of the proposed 2D WFM is below 2.0 keV and therefore complies with the requirements. The only requirement not fully met by the proposed 2D WFM is the energy resolution. The spectral resolution offered by the 2D WFM system is of the order of 600 – 800 eV, which is higher than the required values. However spectral simulations have shown that for the science cases considered today, the spectroscopic advantage of SDD over DSSD, due its higher energy resolution (300 eV vs. 800 eV), is not relevant once the effective area and realistic integration times are considered. Indeed, despite the fact that e.g. absorption edges are in principle better detected by the SDD option, the available statistics for the different science cases presented, shows that all the relevant spectral features are either detected or non-detected by both systems, with only slight advantage for the SDD option in terms of significance. Indeed the advantage (better signal-to-noise ratio) of SDDs for the detection of discrete spectral features is based on one example only of GRB, and on extremely rare events for X-ray bursts. In addition, those differences are relevant only for the triggered observations, since, due to telemetry limitations, the WFM will transmit data binned in only 8 to 16 energy channels making the better spectral capabilities of SDDs with respect to DSSDs not significant.

Table 2. WFM requirement and goal parameters and estimated parameters of the proposed 2D WFM system (1 camera).

| Req. | Parameter | Requirement | Goal | 2DWFM camera |
|---|---|---|---|---|
| **SR 1** | Location Accuracy | < 60" (25 SNR) | < 30" (25 SNR) | < 16" (25 SNR) |
| **SR 2** | Angular Resolution | < 5' | < 3' | 4.12' ± 0.22' |
| **SR 3** | Peak Sensitivity 2-30 keV 50ks 5σ | 5 mCrab | 2 mCrab | 2.06 mCrab |
| SR 4 | Field of View | 1 π sr | 1.5 π sr | 1.9 sr ZRFOV and 0.72 π sr (50% eff) |
| SR 5 | Energy Range | 2 – 50 keV | 1.5 – 50 keV | 1.7-2 – 50 keV |
| SR 6 | Energy Resolution | 500 eV | 300 eV | ≤ 800 eV |
| SR 7 | Time Resolution | 300 s / 10 μs | 150 s / 5 μs | 300 s / 1 μs |

Not surprisingly the 2D imaging system **attains the goal values of nearly all the WFM imaging requirements (SR 1 to SR 4) which are the priority requirements for the WFM**. Note that requirement on PSLE is reached even if no cleaning at all is performed on the images, i.e. assuming that the residual systematic level is the one produced by all the coding noise of the sources in the partially coded FOV. This is a strong characteristic of 2D coded mask systems based on random pattern masks and relatively small distributed mask elements. The angular resolution is intermediate between requirement and goal but could be easily improved increasing the dimensions and weight of the system to reach the goal value of 3'. The key element of the system is that the Point Spread Function is a simple fine 2D Gaussian with nearly totally flat secondary lobes. This is a strong improvement with respect to the PSF of the original 1.5 D design which, basically, is a cross of dimensions 5' × 8° along one axis and 8° × 5' along the other axis, plus a large repetition of such crossed shaped features all over the FOV (secondary lobes).

What allows the 2D system to have such nice properties is that the WFM-2D takes advantage of the in-built very simple photon interaction localization of the DSSD. Each interacting photon producing a signal in one strip (or in two neighboring strips) is localized, at least, with 180 μm accuracy, which is the pitch of the camera. The position is then determined by the hit pixel pattern only, which necessitates no calculation at all. There is no need to treat double or multiple events, the readout can be parameterized in order to reject events with a higher multiplicity than two or not-neighboring doubles, to optimize background rejection or confusing interaction determination. These imaging properties do not limit the real time localization of rapid transient events. Indeed one can show that with a proper design of the mask pattern, the 2D system will be able to detect and localized about 100 GRBs in real time with 4-6 arcmin error radii, and, after refinement, with a ~1 arcmin error radius permitting large robotic telescopes to hunt the counterpart of (possibly cosmological) Gamma-Ray Bursts (GRB).

Such imaging properties are indeed extremely important for the Wide Field Monitor. Specific science cases illustrate the real scientific gain of the 2D WFM configuration.

- The imaging simulations we have performed show that the proposed 2D system will fully disentangle the inner 2° × 2° of the Galaxy. The Galactic Center is the most active region of the sky in terms of X-ray transients and source variability, simply because of the dense concentration of objects. The capability to localize and measure them will add a real important contribution to the WFM observatory science.

- Recent exciting results have been obtained in X-ray and infrared ranges concerning the today-very-quiet Galactic Center super massive black hole Sgr A*. These results show that Sgr A* must have been in the past [24] and will be in the near future [11] much more active than at present times. At the time of the LOFT operations, Sgr A* could be in a much higher levels of activity than shown today and it would be mandatory to observe it. Note that a performing imager able to properly disentangle the different components within the LAD FOV will allow, to a certain extent, to analyze the LAD data even in confused regions like the Galactic Center.

- The very large field of view and the good point spread function will also allow a precise measurement of extended objects such as supernova remnants, pulsar wind nebulae and the Galactic ridge emission.

- Finally, as shown by the imaging simulations, a deep 2-50 keV all sky survey will be readily obtained at the end of the mission thanks to the high quality 2D sky images and the simple summing procedure that the proposed system allows. After 5 years lifetime, sensitivity of the order of 100 μCrab will be achieved up to 10 keV.